\documentclass{article}

\PassOptionsToPackage{numbers}{natbib}

    \usepackage[final]{neurips_2022}

\usepackage[utf8]{inputenc} 
\usepackage[T1]{fontenc}    
\usepackage{hyperref}       
\usepackage{url}            
\usepackage{booktabs}       
\usepackage{amsfonts}       
\usepackage{nicefrac}       
\usepackage{microtype}      
\usepackage{xcolor}         

\usepackage{amsmath}
\usepackage{algorithm2e}

\usepackage{multirow}
\usepackage{graphicx}

\title{Improving the Robustness of DistilHuBERT to Unseen Noisy Conditions via Data Augmentation, Curriculum Learning, and Multi-Task Enhancement }

\author{%
\hspace{-1.2cm}
Heitor Guimarães$^1$ \quad Arthur Pimentel$^1$ \quad Anderson R. Avila$^2$ \quad Mehdi Rezagholizadeh$^2$ \quad Tiago H. Falk$^1$ \\
$^1$Institut National de la Recherche Scientifique \quad $^2$Huawei Noah’s Ark Lab\\
\texttt{\{heitor.guimaraes,arthur.pimentel,tiago.falk\}@inrs.ca}\\
\texttt{\{anderson.avila,mehdi.rezagholizadeh\}@huawei.com}
}

\begin{document}

\maketitle

\begin{abstract}
  Self-supervised speech representation learning aims to extract meaningful factors from the speech signal that can later be used across different downstream tasks, such as speech and/or emotion recognition. Existing models, such as HuBERT, however, can be fairly large thus may not be suitable for edge speech applications. Moreover, realistic applications typically involve speech corrupted by noise and room reverberation, hence models need to provide representations that are robust to such environmental factors. In this study, we build on the so-called DistilHuBERT model, which distils HuBERT to a fraction of its original size, with three modifications, namely: (i) augment the training data with noise and reverberation, while the student model needs to distill the clean representations from the teacher model; (ii) introduce a curriculum learning approach where increasing levels of noise are introduced as the model trains, thus helping with convergence and with the creation of more robust representations; and (iii) introduce a multi-task learning approach where the model also reconstructs the clean waveform jointly with the distillation task, thus also acting as an enhancement step to ensure additional environment robustness to the representation. Experiments on three SUPERB tasks show the advantages of the proposed method not only relative to the original DistilHuBERT, but also to the original HuBERT, thus showing the advantages of the proposed method for ``in the wild'' edge speech applications.
\end{abstract}

\section{Introduction}

Self-supervised learning (SSL) enables the learning of meaningful and disentangled features from unlabeled speech data that can be used across different tasks. These representations are becoming essential and are now part of several state-of-the-art speech applications \cite{baevski2020wav2vec, hsu2021hubert}. Widely-used representations include wav2vec 2.0 \cite{baevski2020wav2vec} and HuBERT \cite{hsu2021hubert}. The latter is an SSL model that extracts features directly from the raw speech signal. Via an offline k-means clustering step, the model learns combined acoustic and language representations from masked inputs using a BERT-like reconstruction loss. 

Existing models, however, have two limitations when edge applications are considered: (1) their large model sizes, and (2) their robustness to environmental conditions not seen during training. For example, pre-trained models can range from 95 million parameters to 2 billion \cite{babu2021xls}, which makes training and deployment on edge devices difficult. To overcome this issue, different model compression schemes have been explored. DistilHuBERT, for example, applied knowledge distillation on the HuBERT model and was able to reduce the number of parameters to a fraction of the original \cite{chang2022distilhubert}. LightHuBERT, in turn, proposed a 2-stage distillation approach to also reduce the size of the original HuBERT \cite{wang2022lighthubert}.

Regarding environmental robustness, commonly with in-the-wild speech applications, there is a shift in the distribution of the test data relative to the distribution of the data used to train the models. This domain shift can be detrimental to downstream task performance. The work in \cite{huang2022improving}, for example, proposes the use of domain adversarial training and data augmentation to improve HuBERT robustness to unseen conditions; such system is termed Robust HuBERT.

In this study, we propose to combine several innovations to make HuBERT not only smaller, but also more robust to unseen environmental conditions. In particular, we propose to combine  data augmentation, curriculum learning, and multi-task training to ensure that the extracted representation is robust. As a proof of concept, we implement these modifications on top of the DistilHuBERT model to benefit from the compression gains already achieved. 





\section{Proposed Methods}


\subsection{Building on DistilHuBERT}
DistilHuBERT uses knowledge distillation on top of the HuBERT model to achieve model compression. In our case, the teacher model is the HuBERT Base model comprised of 12 Transformer layers and 95 million parameters, and the student model has only two Transformer layers and roughly four times fewer parameters. All the parameters are initialized with the values from the teacher model. Distillation occurs with the student model learning a representation which is the input to three prediction heads responsible for predicting the representations from the \{4\textsuperscript{th}, 8\textsuperscript{th}, and 12\textsuperscript{th}\} layers of the teacher model. The prediction heads are discarded after training, making the model efficient and compact. In addition, the distillation mechanism is optimized for a loss function that combines reconstruction elements with a cosine similarity metric, as follows:
\begin{equation} \label{eq:distilhubert_loss}
    \mathcal{L}_{\text{KD}} = \sum_{l \ in \{4, 8, 12\}}\sum_{t=1}^T{\left[ \frac{1}{D}||\pmb{h}^{(l)}_t - \hat{\pmb{h}}^{(l)}_t||_1 - \log{\left(\sigma(\cos( \pmb{h}^{(l)}_t, \hat{\pmb{h}}^{(l)}_t )) \right)} \right]}. 
\end{equation}

The variable $l$ corresponds to the layers from the teacher model to distill, T is the number of time steps, $\pmb{h}^{(l)}_t$ is the $D$-dimensional feature vector extracted from the $l$\textsuperscript{th} layer of the teacher model, $\hat{\pmb{h}}^{(l)}_t$ is the output from the prediction head, and $\sigma(\cdot)$ is the sigmoid function. As DistilHuBERT has shown excellent results compared to the original HuBERT model, it serves as a starting point for our work where the three proposed innovations are implemented on top.

\subsection{Modification \#1: Data augmentation}
Most SSL models rely on learning speech representations from clean readings of audiobook data, such as LibriSpeech \cite{panayotov2015librispeech} or LibriLight \cite{kahn2020libri}. Even though these models can learn essential characteristics from this type of speech, real-world deployment often involves noisy speech utterances which degrade system performance. Data augmentation has commonly been applied to improve the robustness to unseen conditions \cite{huang2022improving}. Here, we propose to perform an online contamination of the data during the distillation process. In particular, the student model receives the noisy data as input, but the network's target is to reconstruct the clean features of the teacher model. At training time, given a batch of clean speech utterances, we uniformly sample one action to be applied to each utterance in the batch: 
\begin{enumerate}
    \item[$(a1)$] No changes are made to the training utterance,
    \item[$(a2)$] Contaminate the utterance with additive noise with signal-to-noise ratio randomly chosen from $[0, 20]$ dB,
    \item[$(a3)$] Convolve the speech waveform with a room impulse response representative of either a small, medium, or a large sized room, and
    \item[$(a4)$] Both $(a2)$ and $(a3)$ actions are jointly applied.
\end{enumerate}

\subsection{Modification \#2: Curriculum/Progressive Learning}
Inspired by the ideas of curriculum learning \cite{bengio2009curriculum}, we propose a progressive training mechanism where the network gradually receives more deteriorated speech samples during training iterations and not randomly from the beginning of the optimization process. Based on the literature, we expect the models trained on this regime to have better generalization capabilities, which can lead us to reinforce the model behavior in learning speech features disentangled from noise. In our work, we gradually increase the difficulty of the online contamination process by defining a custom sampling procedure instead of allowing every scenario from the beginning. This process is done until we reach half of the training steps, and after, the same regime of the previous section is applied. The pseudocode for this step is illustrated in the Supplementary Material Section.

\subsection{Modification \#3: Multi-task enhancement learning}
Speech enhancement is a common processing method used to remove detrimental factors from the speech signal. Motivated by \citet{wang2022improving}, we propose a multi-task learning approach, where beyond learning to reconstruct the teacher's representations, we propose an additional enhancement head responsible to rebuild the clean speech waveform from the learned representation. The goal with this step is to enforce the upstream model to carry enough information about the speech itself and not the noise components. Here, a BiLSTM layer followed by seven transposed convolutions and GELU activation functions is used, as well as the following loss function: $\mathcal{L} = \mathcal{L}_{\text{KD}} + \lambda\mathcal{L}_{\text{enh}},$ where $\mathcal{L}_{\text{enh}}$ is the enhancement loss and $\lambda \in \mathbb{R}$ is a hyperparameter that controls the importance of the enhancement step on the total loss. Notice that multiple choices can be made about what enhancement loss to use. Here, two enhancement losses are explored: (i) an L1-based reconstruction loss directly on the waveform samples and (ii) an L1-based reconstruction loss applied to the spectral magnitude of the reconstructed signal. These are denote by $\mathcal{L}_{\text{L1-wav}}$ and $\mathcal{L}_{\text{L1-freq}}$, respectively.

\vspace{-2mm}
\section{Experimental Setup}

\subsection{Datasets}
The training data used is the LibriSpeech dataset \cite{panayotov2015librispeech}, a corpus with 960 hours of audiobook readings with a 16 kHz sampling rate and 16-bit resolution. The training data consists of a subset of 460 hours of clean audio in a studio-like condition, while the other 500 hours do not have the same quality standard. To augment the training set, two additional noise datasets are used, namely MUSAN \cite{musan2015}, and UrbanSound8K \cite{Salamon:UrbanSound:ACMMM:14}, as well as the OpenSLR28 dataset \cite{ko2017study} of room impulse responses (RIR). 

The MUSAN dataset contains 6 hours of noise recordings (16 kHz sample rate) in a wide variety of categories, including office-like noises, babble noise, and natural sounds, such as wind and animals, to name a few. The UrbanSound8K dataset contains approximately 9 hours of audio distributed across ten labels: air conditioner, car horn, children playing, dog bark, drilling, engine idling, gunshot, jackhammer, siren, and street music. We removed the children playing and street music conditions to focus on non-speech like noise sources in this first analysis. The dataset has a sample rate of 44.1 kHz, which was resampled to 16 kHz. Lastly, the openSLR28 dataset contains 325 RIRs from multiple sources, such as small meeting rooms and large churches with up to 570 m\textsuperscript{2}.

At test time, it is important to test the model with unseen conditions. To this end, we rely on two separate datasets to corrupt the test set. The first is the Deep Noise Suppression 4 (DNS4) challenge dataset \cite{dubey2022icassp}, comprised of approximately 180 hours of noise extracted from 150 categories with no presence of speech. For reverberation, we used the OpenSLR26 dataset \cite{ko2017study}, which contains 60,000 simulated room impulse responses corresponding to various small-, medium-, and large-sized rooms.

\subsection{Pre-training}
To gauge the benefits of the different proposed modifications, four different combinations are tested. First, denoted as experiment A, explores the impact of only applying data augmentation on top of DistilHuBERT. Next, experiment B  adds the progressive learning approach. Lastly, experiments  C1 and C2 explore the further addition of the enhancement head where the loss functions $\mathcal{L}_{\text{L1-wav}}$ and $\mathcal{L}_{\text{L1-freq}}$ are used, respectively. We set $\lambda = 10$ and $\lambda = 1$ for experiments C1 and C2, respectively. 

In all cases, the training recipe described in \cite{chang2022distilhubert} is followed. Upstream models are trained using a single NVidia A100 GPU. Experiments A and B take approximately 30 hours to train, whereas experiments C1 and C2 took each roughly 43 hours to train. We use the AdamW optimizer, with a batch of 24 utterances, for 200k iterations, whereas after 14k updates, the learning rate linearly decays from $2\times10^{-4}$ to zero. As benchmarks, we use the original HuBERT Base model, its Robust version, and two compressed models: LightHuBERT and DistilHubert.

\subsection{Downstream tasks}
A subset of three downstream tasks from the SUPERB benchmark \cite{yang2021superb} are chosen to evaluate the robustness of the proposed modifications, namely: keyword spotting, intent classification, and emotion recognition. For all downstream tasks, the evaluation metric is the standard accuracy score; thus, higher values are better. To test the robustness of the methods and models, four evaluation scenarios are considered: {clean} $(c)$, {noise-only} $(n)$, {reverberation-only} $(r)$, and {noise-plus-reverberation} $(n+r)$. The test set meets the same criteria as SUPERB downstream tasks in the clean setting. For the noisy test conditions, in turn, additive noise with signal-to-noise ratios ranging from $[0, 20]$ dB are added. For the reverberation condition, a random room impulse response is uniformly sampled and applied to the test set utterance. Lastly, for the noise-plus-reverberation condition, both noise and reverberation are jointly applied to the test set, thus representing the most challenging scenario of being in a noisy room. A custom seed is applied to ensure all models are evaluated with the same degradations.

\vspace{-2mm}
\section{Results and Discussion}
\begin{table}
    \centering
    \caption{Experimental results for keyword spotting, intent classification, and emotion recognition under clean (\textbf{c}), noisy (\textbf{n}), reverberation (\textbf{r}), and noise-plus-reveberation (\textbf{n+r}) test conditions.}
    \label{tab:results}
    \resizebox{\textwidth}{!}{%
    \begin{tabular}{cccccccccccccc}
        \toprule
        & & \multicolumn{4}{c}{Keyword Spotting} & \multicolumn{4}{c}{Intent Classification} & \multicolumn{4}{c}{Emotion Recognition} \\

        \cmidrule(lr){3-6}
        \cmidrule(lr){7-10}
        \cmidrule(lr){11-14}

        Upstream & \#params (M) & (c) & (n) & (r) & (n + r) & (c) & (n) & (r) & (n + r) & (c) & (n) & (r) & (n + r)  \\ 

        \midrule

        HuBERT Base (Teacher) & 95 & 96.30 & 84.10 & 61.70 & 53.49 & 98.34 & 80.78 & 75.77 & 56.60 & 64.75 & 52.57 & 40.72 & 32.06 \\
        Robust HuBERT & 95 & 96.33 & 91.46 & 74.81 & 66.80 & 98.66 & 92.75 & 85.32 & 71.53 & 64.59 & 57.83 & 41.44 & 34.83 \\
        LightHuBERT Small & 27 & 96.11 & 82.80 & 72.35 & 57.12 & 98.60 & 80.81 & 88.19 & 65.20 & 63.88 & 48.65 & 35.51 & 31.24 \\
        DistilHuBERT & 24 & 96.14 & 83.45 & 59.20 & 52.26 & 94.12 & 54.23 & 47.72 & 28.37 & 62.55 & 48.67 & 31.35 & 28.30\\
        \textbf{Proposed (Exp. A)} & 24 & 96.59 & 92.31 & 85.98 & 79.52 & 94.15 & 81.68 & 83.42 & 67.23 & 61.71 & 50.89 & 40.83 & 33.95 \\
        \textbf{Proposed (Exp. B)} & 24 & 96.56 & 92.05 & 86.98 & 80.30 & 94.20 & 83.07 & 83.81 & 67.39 & 62.42 & 52.52 & 39.12 & 32.39 \\
        \textbf{Proposed (Exp. C1)} & 24 & 96.69 & 91.40 & 85.98 & 79.65 & 94.57 & 85.50 & 82.76 & 70.29 & 61.88 & 55.02 & 39.27 & 32.17 \\
        \textbf{Proposed (Exp. C2)} & 24 & 96.92 & 91.85 & 85.30 & 79.00 & 93.91 & 83.39 & 81.18 & 64.70 & 62.96 & 51.48 & 39.24 & 32.78 \\
        \midrule[\heavyrulewidth]
        \bottomrule
    \end{tabular}
    }
\end{table}

Table~\ref{tab:results} presents the experimental results for the benchmarks and proposed methods on the three downstream tasks. As can be seen, the proposed innovations outperformed the original DistilHuBERT model for all of the three SUPERB tasks, even in clean conditions. In fact, just adding data augmentation (Expt. A) improved clean speech accuracy for the keyword spotting and intent classification tasks, thus showing the benefits of data augmentation. Progressive learning, in turn, (Expt. B) showed to be particularly important for the noisy/reverberant conditions for keyword spotting and intent classification. The additional enhancement step also showed to provide additional gains, with the waveform-based loss outperforming the magnitude spectral one. 

In fact, for keyword spotting in noise, reverberation, and noise-plus-reverberation conditions, the proposed methods with only 24M parameters outperformed the larger teacher and Robust HuBERT models, each comprised of 95M parameters. For the two other tasks under unseen conditions, the proposed methods outperformed the original teacher model and obtained results in line with those of Robust HuBERT, whilst requiring roughly one quarter of the number of parameters. The proposed multi-task speech enhancement step showed to be particularly useful for when additive noise was present, especially for intent classification and emotion recognition. In the future, other perceptual losses will be explored to see if further improvements can be obtained under reverberant conditions. 



\section{Conclusions}
In this work, we propose three modifications to be implemented on top of DistilHubert. The aim is to develop a cross-task speech representation that is compact and, at the same time, robust to unseen noisy conditions. The modifications include addition of data augmentation, progressive learning, and multi-task learning where speech enhancement is jointly performed. Experiments on three SUPERB tasks, namely keyword spotting, intent classification, and emotion recognition, show the proposed model outperforming the original DistilHuBERT across all tested noisy conditions, sometimes even outperforming the original teacher HuBERT model and its robust variant. The obtained results are promising and suggest that the proposed method can be useful for future edge speech applications.



\bibliography{refs}

\begin{thebibliography}{15}
\providecommand{\natexlab}[1]{#1}
\providecommand{\url}[1]{\texttt{#1}}
\expandafter\ifx\csname urlstyle\endcsname\relax
  \providecommand{\doi}[1]{doi: #1}\else
  \providecommand{\doi}{doi: \begingroup \urlstyle{rm}\Url}\fi

\bibitem[Babu et~al.(2021)Babu, Wang, Tjandra, Lakhotia, Xu, Goyal, Singh, von
  Platen, Saraf, Pino, et~al.]{babu2021xls}
A.~Babu, C.~Wang, A.~Tjandra, K.~Lakhotia, Q.~Xu, N.~Goyal, K.~Singh, P.~von
  Platen, Y.~Saraf, J.~Pino, et~al.
\newblock Xls-r: Self-supervised cross-lingual speech representation learning
  at scale.
\newblock \emph{arXiv preprint arXiv:2111.09296}, 2021.

\bibitem[Baevski et~al.(2020)Baevski, Zhou, Mohamed, and
  Auli]{baevski2020wav2vec}
A.~Baevski, Y.~Zhou, A.~Mohamed, and M.~Auli.
\newblock wav2vec 2.0: A framework for self-supervised learning of speech
  representations.
\newblock \emph{Advances in Neural Information Processing Systems},
  33:\penalty0 12449--12460, 2020.

\bibitem[Bengio et~al.(2009)Bengio, Louradour, Collobert, and
  Weston]{bengio2009curriculum}
Y.~Bengio, J.~Louradour, R.~Collobert, and J.~Weston.
\newblock Curriculum learning.
\newblock In \emph{Proceedings of the 26th annual international conference on
  machine learning}, pages 41--48, 2009.

\bibitem[Chang et~al.(2022)Chang, Yang, and Lee]{chang2022distilhubert}
H.-J. Chang, S.-w. Yang, and H.-y. Lee.
\newblock Distilhubert: Speech representation learning by layer-wise
  distillation of hidden-unit bert.
\newblock In \emph{ICASSP 2022-2022 IEEE International Conference on Acoustics,
  Speech and Signal Processing (ICASSP)}, pages 7087--7091. IEEE, 2022.

\bibitem[Dubey et~al.(2022)Dubey, Gopal, et~al.]{dubey2022icassp}
H.~Dubey, V.~Gopal, et~al.
\newblock Icassp 2022 deep noise suppression challenge.
\newblock In \emph{Proc. ICASSP}, 2022.

\bibitem[Hsu et~al.(2021)Hsu, Bolte, Tsai, Lakhotia, Salakhutdinov, and
  Mohamed]{hsu2021hubert}
W.-N. Hsu, B.~Bolte, Y.-H.~H. Tsai, K.~Lakhotia, R.~Salakhutdinov, and
  A.~Mohamed.
\newblock Hu{BERT}: Self-supervised speech representation learning by masked
  prediction of hidden units.
\newblock \emph{IEEE/ACM Transactions on Audio, Speech, and Language
  Processing}, 29:\penalty0 3451--3460, 2021.

\bibitem[Huang et~al.(2022)Huang, Fu, Zhang, and Lee]{huang2022improving}
K.~P. Huang, Y.-K. Fu, Y.~Zhang, and H.-y. Lee.
\newblock Improving distortion robustness of self-supervised speech processing
  tasks with domain adaptation.
\newblock \emph{arXiv:2203.16104}, 2022.

\bibitem[Kahn et~al.(2020)Kahn, Rivi{\`e}re, et~al.]{kahn2020libri}
J.~Kahn, M.~Rivi{\`e}re, et~al.
\newblock Libri-light: A benchmark for {ASR} with limited or no supervision.
\newblock In \emph{Proc. IEEE ICASSP}, pages 7669--7673, 2020.

\bibitem[Ko et~al.(2017)Ko, Peddinti, Povey, Seltzer, and
  Khudanpur]{ko2017study}
T.~Ko, V.~Peddinti, D.~Povey, M.~L. Seltzer, and S.~Khudanpur.
\newblock A study on data augmentation of reverberant speech for robust speech
  recognition.
\newblock In \emph{Proc. IEEE ICASSP}, pages 5220--5224, 2017.

\bibitem[Panayotov et~al.(2015)Panayotov, Chen, Povey, and
  Khudanpur]{panayotov2015librispeech}
V.~Panayotov, G.~Chen, D.~Povey, and S.~Khudanpur.
\newblock Librispeech: an asr corpus based on public domain audio books.
\newblock In \emph{2015 IEEE international conference on acoustics, speech and
  signal processing (ICASSP)}, pages 5206--5210. IEEE, 2015.

\bibitem[Salamon et~al.(2014)Salamon, Jacoby, and
  Bello]{Salamon:UrbanSound:ACMMM:14}
J.~Salamon, C.~Jacoby, and J.~P. Bello.
\newblock A dataset and taxonomy for urban sound research.
\newblock In \emph{22nd {ACM} International Conference on Multimedia
  (ACM-MM'14)}, pages 1041--1044, Orlando, FL, USA, Nov. 2014.

\bibitem[Snyder et~al.(2015)Snyder, Chen, and Povey]{musan2015}
D.~Snyder, G.~Chen, and D.~Povey.
\newblock {MUSAN}: {A} {M}usic, {S}peech, and {N}oise {C}orpus, 2015.

\bibitem[Wang et~al.(2022{\natexlab{a}})Wang, Qian, Wang, Wang, Wang, Liu,
  Yoshioka, Li, and Wang]{wang2022improving}
H.~Wang, Y.~Qian, X.~Wang, Y.~Wang, C.~Wang, S.~Liu, T.~Yoshioka, J.~Li, and
  D.~Wang.
\newblock Improving noise robustness of contrastive speech representation
  learning with speech reconstruction.
\newblock In \emph{ICASSP 2022-2022 IEEE International Conference on Acoustics,
  Speech and Signal Processing (ICASSP)}, pages 6062--6066. IEEE,
  2022{\natexlab{a}}.

\bibitem[Wang et~al.(2022{\natexlab{b}})Wang, Bai, Ao, Zhou, Xiong, Wei, Zhang,
  Ko, and Li]{wang2022lighthubert}
R.~Wang, Q.~Bai, J.~Ao, L.~Zhou, Z.~Xiong, Z.~Wei, Y.~Zhang, T.~Ko, and H.~Li.
\newblock Lighthubert: Lightweight and configurable speech representation
  learning with once-for-all hidden-unit bert.
\newblock \emph{arXiv preprint arXiv:2203.15610}, 2022{\natexlab{b}}.

\bibitem[Yang et~al.(2021)Yang, Chi, et~al.]{yang2021superb}
S.-w. Yang, P.-H. Chi, et~al.
\newblock Superb: Speech processing universal performance benchmark.
\newblock \emph{arXiv preprint arXiv:2105.01051}, 2021.

\end{thebibliography}

\newpage


\section{Supplementary Material}
\subsection*{Pseudocode for progressive learning experiment} \label{appx:algorithm}

Algorithm~\ref{alg:inc_sampling} shows the pseudocode to describe the progressive learning strategy adopted in this work. It is a generalization of the data augmentation strategy described in experiment A. Notice that, if $\tau = 0$ and $t = 1$, we recover the augmentation stage without curriculum learning.

\RestyleAlgo{ruled}
\LinesNumbered
\SetKwComment{Comment}{/* }{ */}

\begin{algorithm}[hbt!]
\caption{Sample strategy for SNR-Progressive Learning}\label{alg:inc_sampling}
\textbf{given} a batch of utterances $U = \{u_i\}_{i=0}^L$, a set of noise files $U_n$, a set of room impulse responses $U_r$, the current iteration $it$, and the total number of iterations $N$ \;
\For{each utterance $u_i$}{
a $\gets$ sample an action from the equiprobable sample space $\{a1,~a2,~a3,~a4\}$ \;
    \uIf{$a == a2$}{
    $\tau \gets 20\left(1 - \frac{2it}{N}\right)$ \textbf{if} $it \leq N/2$ \textbf{else} 0 \Comment*[r]{Lower-bound for the SNR value}
    $snr \gets$ sample an integer number from a uniform distribution $\mathcal{U}(\tau, 20)$ \;
    $p_n \gets$ sample a random number from a uniform distribution $\mathcal{U}(0, 1)$ \;
    $u_n$ $\gets$ sample from $U_n$ \textbf{if} $p_n \leq 0.7$ \textbf{else} return a narrow-band white noise $\mathcal{N}(0, 1)$ \;
    $u_i \gets$ applyNoise($u_i$, $u_n$, $snr$)
    
  }
  \uElseIf{$a == a3$}{
    $t \gets \frac{2it}{N}$ \textbf{if} $it \leq N/2$ \textbf{else} 1 \Comment*[r]{Threshold for applying reverberation}
    $p_r \gets$ sample a random number from a uniform distribution $\mathcal{U}(0, 1)$ \;
    \If{$p_r \leq t$}{
        $u_r \gets$ uniformly sample a RIR from $U_r$ \;
        $u_i \gets$ applyReverberation($u_i$, $u_r$) \;
    }
  }
  \uElseIf{$a == a4$}{
    $u_i \gets$ apply the same mechanism from the action $a2$ \;
    $u_i \gets$ apply the same mechanism from the action $a3$ \;
  }
  \Else{
    \textit{Do nothing} \;
  }
}
\end{algorithm}

\end{document}